\documentclass[%
 reprint,
reprint,onecolumn,11pt,
amsmath,amssymb,aps,
prb,
]{revtex4-1}

\usepackage{verbatim}
\usepackage{graphicx}
\normalfont \topmargin -0.5 cm
\begin{document}

\title{Quadrupolar induced suppression of nuclear spin bath fluctuations in self-assembled quantum dots}

\author{E.~A.~Chekhovich$^1$}
\author{M.~Hopkinson$^2$}
\author{M.~S.~Skolnick$^1$}
\author{A.~I.~Tartakovskii$^1$}
\affiliation{$^1$Department of Physics and Astronomy, University
of Sheffield, Sheffield S3 7RH, UK} \affiliation{$^2$Department of
Electronic and Electrical Engineering, University of Sheffield,
Sheffield S1 3JD, UK}

\date{\today}


\maketitle


\textbf{Decoherence in quantum logic gates (qubits) due to
interaction with the surrounding environment is a major obstacle
to the practical realization of quantum information technologies.
For solid state electron-spin qubits the interaction with nuclear
spins is the main problem
\cite{RMPReview,Bluhm,DeGreveHole,Koppens2,HoleNW,GreilichHoleQbit,Kuhlmann}.
One particular, ineradicable source of electron decoherence arises
from decoherence of the nuclear spin bath, driven by
nuclear-nuclear dipolar interactions. Due to its many-body nature
nuclear decoherence is difficult to predict
\cite{DasSarma2003,Sham2006}, especially for an important class of
strained nanostructures where nuclear quadrupolar effects have a
significant but largely unknown impact \cite{Sinitsyn2012}. Here
we report direct measurement of nuclear spin bath coherence in
individual strained InGaAs/GaAs quantum dots: nuclear spin-echo
coherence times in the range $T_2\approx1.2 - 4.5$~ms are found.
Based on these $T_2$ values we demonstrate that quadrupolar
interactions make nuclear fluctuations in strained quantum dots
much slower compared to lattice matched GaAs/AlGaAs structures
\cite{Bluhm}. Such fluctuation suppression is particularly strong
for arsenic nuclei due to the effect of atomic disorder of gallium
and indium alloying. Our findings demonstrate that quadrupolar
effects can help to solve the long-standing challenge of designing
a scalable hardware for quantum computation: III-V semiconductor
spin-qubits can be engineered to have a noise-free nuclear spin
bath (previously achievable only in nuclear spin-0 semiconductors
\cite{Saeedi15112013,CNTQbit,NeumannDiamond}, where qubit network
interconnection and scaling is challenging).}

Quantum dots in III-V semiconductors have many favourable
properties for applications in quantum information processing
including strong interaction with light offering excellent optical
interfacing, manipulation at ultrafast speeds and advanced
manufacturing technology \cite{Koppens2,HoleNW,Reilly,Bluhm}.
However, all atoms of groups III and V have nonzero nuclear
magnetic moments. Thus instead of an ideal two-level quantum
system, the spin of a single electron in a quantum dot is
described by the so called ''central spin'' problem
\cite{Merkulov,DasSarma2003,Sham2006}, where the electron
(central) spin is subject to magnetic interaction with an ensemble
of $10^4-10^6$ nuclear spins. This hyperfine interaction results
in decoherence, i.e. decay of the phase information encoded in
electron spin \cite{Merkulov,DasSarma2003,Sham2006,Bluhm}.

Hyperfine induced decoherence can be greatly reduced by applying
static magnetic field and refocusing control pulses inducing
electron spin echo. With this technique very long electron qubit
coherence times of $\sim200$~$\mu$s were demonstrated in lattice
matched GaAs/AlGaAs quantum dots \cite{Bluhm}. However, the effect
of nuclei can not be eliminated completely due to presence of
nuclear-nuclear (dipole-dipole) magnetic interactions, which cause
spin exchange flip-flops of nuclei, i.e. nuclear spin bath
decoherence. Due to the many-body nature of the nuclear spin bath
such flip-flops induce quasi-random fluctuating magnetic fields
acting on electron spin and causing its decoherence (''spectral
diffusion'' process \cite{DasSarma2003,DasSarma2006,Sham2006}). It
is thus evident that understanding the nuclear spin coherence is
crucial for predicting the coherence properties of the central
spin. Furthermore, strained quantum dots exhibit large nuclear
quadrupolar interactions (QI) \cite{QNMRArxiv}. It is predicted
that the QI can suppress the nuclear flip-flops resulting in
extended electron spin coherence \cite{KorenevQ}. However, this
possibility is little explored \cite{Sinitsyn2012}, mainly due to
the lack of reliable data on nuclear spin coherence in
self-assembled dots.

Here we develop experimental techniques enabling pulsed nuclear
magnetic resonance (NMR) of as few as $10^4-10^5$ quadrupolar
spins in individual strained InGaAs/GaAs quantum dots. We probe
nuclear coherence by measuring the spin-echo decay times $T_2$,
found to range from $\sim$1.2~ms [for $^{71}$Ga] to 4.5~ms [for
$^{75}$As], which is a factor of $\sim5$ longer compared to
unstrained GaAs/AlGaAs structures -- a direct evidence of the
nuclear spin flip-flop suppression. We then show that the nuclear
flip-flop times $T_{2,ff}$ (relevant for electron spin
decoherence) differ from spin-echo $T_2$ times, but can be
estimated using a first-principle model \cite{Haase1993Echo}. For
gallium and indium the strain-induced inhomogeneous QI results in
$T_{2,ff}\sim5$~ms, a factor of $\sim$3 - 8 increase compared to
lattice-matched structures. By contrast, for arsenic a much
stronger flip-flop suppression is found ($T_{2,ff}\gg$5~ms), an
effect explained by additional inhomogeneous QI arising from
random alloy mixing of gallium and indium atoms
\cite{KnijnIntermixing}. Such atomic-scale disorder opens a new
prospect for using the excellent properties of III-V quantum dots
to build nuclear-spin-noise free solid-state qubits: This can now
be done without resorting to materials with zero nuclear spin
(e.g. isotopically pure $^{28}$Si and $^{12}$C)
\cite{Saeedi15112013,CNTQbit,NeumannDiamond}, which have inferior
optical properties, hampering on-chip integration of a large
number of qubits.

Our experiments were performed on individual neutral quantum dots
in InGaAs/GaAs samples, grown by strain-driven self-assembly using
molecular beam epitaxy. The sample was placed in an optical
helium-bath cryostat ($T$=4.2~K). Magnetic field $B_z$ up to 8~T
was applied parallel to the sample growth axis ($Oz$) and light
propagation direction (Faraday geometry). The structures were
investigated using optically detected nuclear magnetic resonance
(ODNMR) techniques which extend the concepts reported in our
recent work \cite{QNMRArxiv,HoleNucIso}. Radio-frequency (rf)
fields $B_{rf}$ perpendicular to $B_z$ are induced by a minicoil
wound around the sample (see further details in Methods and
Supplementary Sections S1 and S2).

\begin{figure}
\includegraphics[bb=22pt 21pt 274pt 260pt]{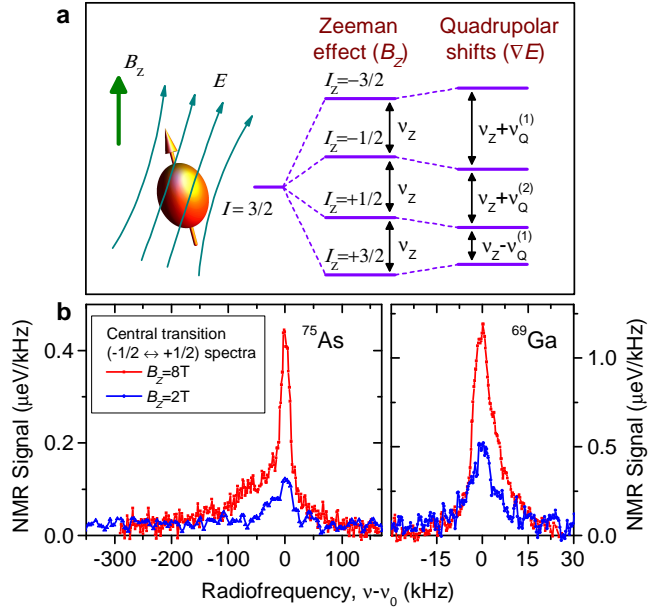}
\caption{\label{fig:NMRSpec}\textbf{Nuclear quadrupolar effects in
strained quantum dots.} \textbf{a,} Schematic representation of a
half-integer quadrupolar nuclear spin ($I=3/2$ is used as an
example) and its energy spectrum. Magnetic field $B_z$ splits the
spin state into a Zeeman ladder, with dipole-allowed NMR
transitions $I_z\leftrightarrow I_z\pm1$ occurring at the same
Larmor frequency $\nu_Z$. Due to nonzero quadrupolar moment the
electric field gradient $\nabla \vec{E}$ (induced e.g. by strain)
causes first and second order NMR frequency shifts $\nu_Q^{(1)}$,
$\nu_Q^{(2)}$ with $\nu_Q^{(1)}\gg\nu_Q^{(2)}$. \textbf{b,}
Central transition spectra of $^{69}$Ga and $^{75}$As measured in
an InGaAs QD using continuous wave ''inverse'' techniques
\cite{QNMRArxiv} at $B_z=$2 and 8~T. At lower magnetic field the
resonance peaks become weaker and broader, confirming that the
linewidth is determined by the second order shifts
$\nu_Q^{(2)}\propto1/B_z$.}
\end{figure}

In this work we study the four most abundant isotopes: $^{69}$Ga,
$^{71}$Ga, $^{75}$As (spin $I=3/2$) and $^{115}$In (spin $I=9/2$),
all possessing non-zero quadrupolar moments. The proportion of
Ga/In in our dots is estimated as 0.76/0.24 (Ref.
\cite{QNMRArxiv}). The energy level diagram of a quadrupolar
nuclear spin is shown schematically in Fig. \ref{fig:NMRSpec}(a)
for the case of $I=3/2$. Magnetic field $B_z$ induces shifts
proportional to $\propto I_z$, so that all dipole-allowed NMR
transitions ($\Delta I_z=\pm1$) appear at the same frequency
$\nu_Z$. Electric field gradients $\nabla \vec{E}$ induce further
shifts $\propto I_z^2$ to first order of perturbation
\cite{AbrahamBook}. The resulting NMR frequency shifts
$\nu_Q^{(1)}$ are strongly inhomogeneous and are on the order of
few MHz in InGaAs dots \cite{QNMRArxiv}. The central transition
(CT) $-1/2 \leftrightarrow +1/2$ is an exception, since it is
affected by QI only to second order resulting in much smaller
shifts $\nu_Q^{(2)}$ on the order of tens to hundreds of kHz
\cite{AbrahamBook}. The relatively small linewidths greatly
simplify the experiments; thus in what follows we focus on
spectroscopy of CTs only. In particular, selective pulsed NMR of
CTs can be conveniently implemented by choosing the rf amplitude
$B_{rf}$ so that $\nu_Q^{(2)}\lesssim \gamma
B_{rf}/(2\pi)\ll\nu_Q^{(1)}$ ($\gamma$ is the nuclear gyromagnetic
ratio).

Fig. \ref{fig:NMRSpec}(b) shows CT spectra of $^{75}$As and
$^{69}$Ga measured using continuous-wave ''inverse'' NMR
techniques \cite{QNMRArxiv}. At high field $B_z=8$~T the $^{69}$Ga
resonance consists of a single narrow line (FWHM $\sim$9~kHz). The
arsenic resonance consists of a narrow line (FWHM $\sim$30~kHz)
and additional asymmetric ''wings'' stretching up to approximately
-150/+50 kHz. When magnetic field is reduced down to 2~T both
resonances broaden and diminish in amplitude, as expected for a
lineshape determined by second order quadrupolar shifts
$\nu_Q^{(2)}$ (Ref. \cite{AbrahamBook}). The significantly larger
broadening of the $^{75}$As CT resonance is attributed to random
intermixing of the group-III Ga and In atoms creating additional
low-symmetry electric field gradients at arsenic sites
\cite{KnijnIntermixing,QNMRArxiv}. As we demonstrate below such
random quadrupolar shifts result in pronounced suppression of
dipolar nuclear flip-flops and extension of nuclear spin coherence
times.

\begin{figure*}
\includegraphics[bb=47pt 53pt 553pt 272pt]{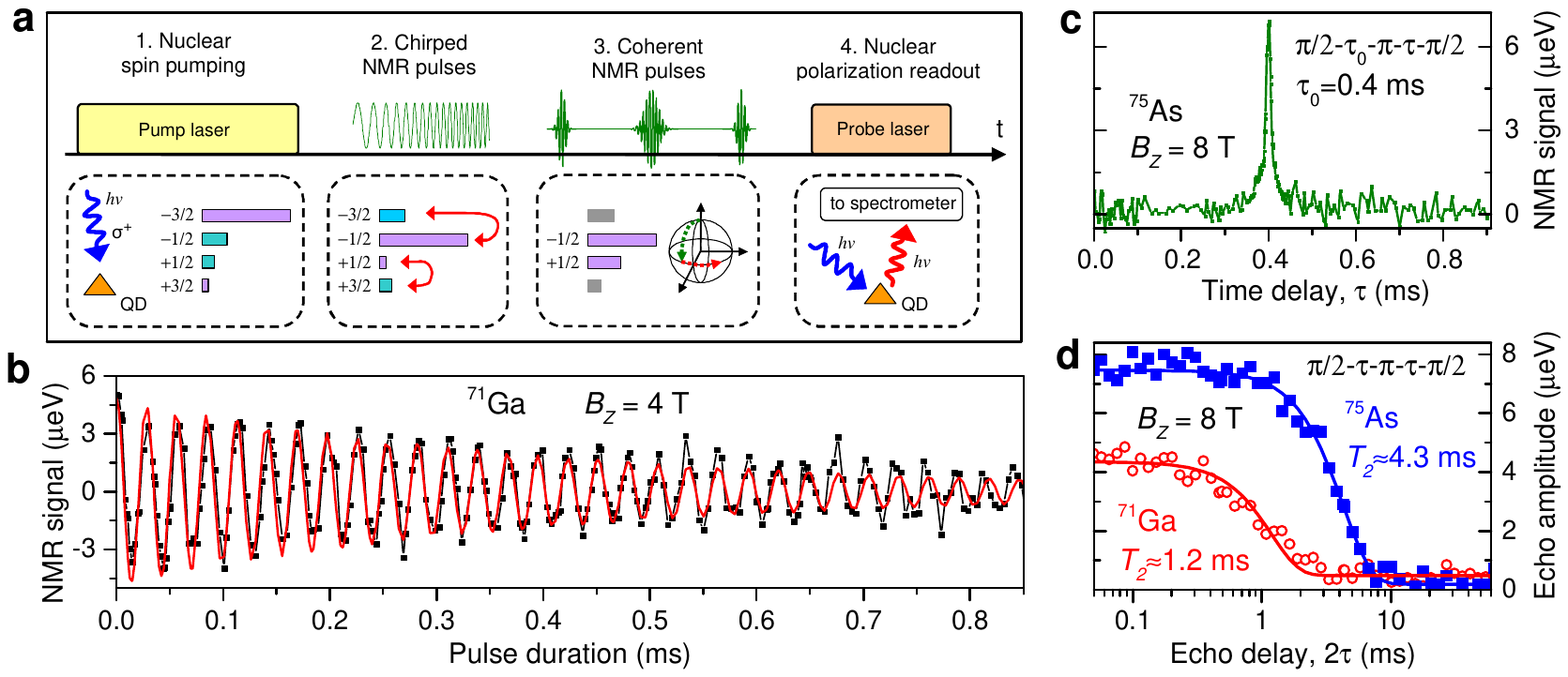}
\caption{\label{fig:CTScheme} \textbf{Pulsed NMR on central
transitions ($-1/2 \leftrightarrow +1/2$) of quadrupolar nuclear
spins in InGaAs quantum dots.} \textbf{a,} Experimental cycle
timing consisting of four stages: (1) Optical pumping of a large
($>$50\%) nuclear spin polarization \cite{GammonPRL,HoleNucIso},
(2) enhancement of the CT signal by chirped radiofrequency pulses
via population transfer \cite{Haase1993Enhance}, (3) an arbitrary
sequence of radiofrequency pulses selectively exciting the CT, (4)
optical readout of the final nuclear spin polarization using
hyperfine shifts in a photoluminescence spectrum
\cite{GammonPRL,HoleNucIso}. \textbf{b,} Rabi oscillations
measurement (a single rf pulse of a variable duration).
\textbf{c,} Hahn-echo measurement ($\pi/2-\tau_0-\pi-\tau-\pi/2$
sequence with $\tau_0=$0.4~ms): a pronounced spin echo signal is
observed at $\tau=0.4$~ms \textbf{d,} Echo decay measurements at
$B_z=8$~T on $^{71}$Ga (circles) and $^{75}$As (squares): Spin
echo amplitude is plotted as a function of the total delay $2\tau$
of the $\pi/2-\tau-\pi-\tau-\pi/2$ sequence. Lines show Gaussian
decay fitting $\propto \exp(-(2\tau)^2/T_2^2)$ with decay times
$T_2$ characterizing the nuclear spin coherence.}
\end{figure*}

The quadrupolar broadening of NMR spectra in Fig.
\ref{fig:NMRSpec}(b) is inhomogeneous in character. It obscures
the much weaker homogeneous broadening induced by the
nuclear-nuclear interactions which determine the nuclear spin
coherence. In order to access the nuclear spin coherence we use
time-domain (pulsed) NMR \cite{AbrahamBook}: the timing diagram of
the pulsed NMR experiment is shown in Fig. \ref{fig:CTScheme}(a)
(further details on techniques can be found in Methods and
Supplementary Section S2). We start with a Rabi nutation
experiment where a single rf pulse of a variable duration $\tau$
is applied \cite{MakhoninNatMat}. Rabi oscillations of nuclear
polarization are clearly seen in Fig. \ref{fig:CTScheme}(b)
enabling the calibration of $\pi/2$ and $\pi$ rotation pulses. The
decay of Rabi oscillations is due to dephasing caused by
inhomogeneous spectral broadening [Fig. \ref{fig:NMRSpec}(b)].
Such dephasing can be reversed using the Hahn echo sequence
$\pi/2-\tau_0-\pi-\tau-\pi/2$. The result of a measurement with a
fixed delay $\tau_0=0.4$~ms and a variable $\tau$ are shown in
Fig. \ref{fig:CTScheme}(c) where as expected a pronounced spin
echo is observbed at $\tau=\tau_0$.

\begin{figure}
\includegraphics[bb=51pt 49pt 293pt 255pt]{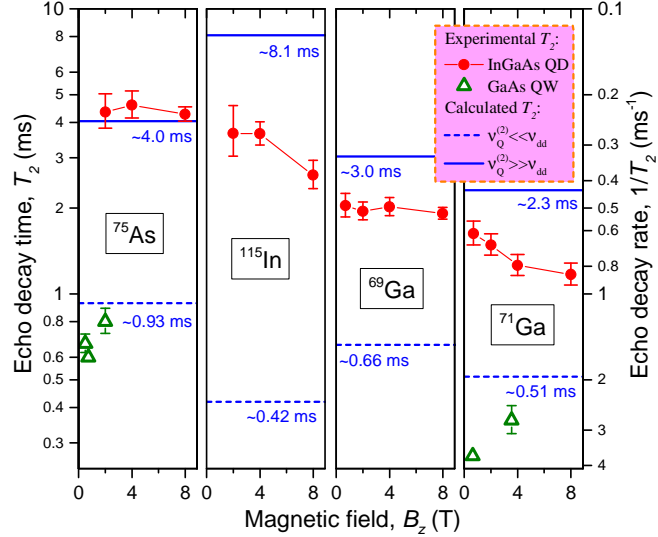}
\caption{\label{fig:CTEcho}\textbf{Nuclear spin-echo decay of
central transitions.} Echo decay times $T_2$ (left scale) and
corresponding decay rates $1/T_2$ (right scale) for four different
isotopes as a function of $B_z$. Circles -- experiment on strained
InGaAs QDs (bars show 90\% confidence intervals), triangles -
measurements on lattice-matched GaAs/AlGaAs quantum wells (QWs)
and QDs (data taken from Refs.
\cite{KondoGaAsEcho,OnoGaAsEcho,IshiharaGaAsEcho,Sanada,MakhoninNatMat}).
Calculated $T_2$ values \cite{Haase1993Echo} are shown for the
case of negligible second order quadrupolar shifts
$\nu_Q^{(2)}\ll\nu_{dd}$ as in lattice matched GaAs/AlGaAs
structures (dashed lines), and for the case of large inhomogeneous
second order quadrupolar shifts $\nu_Q^{(2)}\gg\nu_{dd}$ resulting
in complete suppression of nuclear flip-flops (solid lines).}
\end{figure}

We then turn to the spin-echo decay measurements
($\pi/2-\tau-\pi-\tau-\pi/2$ pulse sequence) where the evolution
times $\tau$ before and after the $\pi$ refocusing pulse are
varied simultaneously. Figure \ref{fig:CTScheme}(d) shows
experimentally measured nuclear spin-echo amplitudes (symbols) as
a function of the total delay time $2\tau$ for $^{71}$Ga and
$^{75}$As isotopes at $B_z=8$~T. Experimental curves are well
fitted by a Gaussian decay function (solid lines) with
characteristic 1/$e$ decay time $T_2\approx$1.18~ms for $^{71}$Ga
and $T_2\approx$4.27~ms for $^{75}$As. The spin-echo sequence
removes the effect of inhomogeneous spectral broadening, with the
echo decay caused solely by the nuclear-nuclear dipolar
interactions \cite{AbrahamBook}: $T_2$ thus characterizes the
coherence of the nuclear spin bath. We have repeated spin-echo
measurements for all four studied isotopes at different magnetic
fields $B_z$. The resulting coherence times $T_2$ (and
corresponding decay rates $1/T_2$) are plotted in Fig.
\ref{fig:CTEcho} by the circles. In addition we have verified the
reproducibility of our results by measuring $T_2$ of $^{75}$As for
another six individual dots from the same sample (see
Supplementary Sec. S3).

In order to examine the effect of QI on the nuclear spin bath
coherence we first compare our experimental $T_2$ times with
previous nuclear spin echo measurements on GaAs/AlGaAs quantum
wells and dots. The data available for $^{75}$As (selective echo
on CT in quantum wells (QWs)
\cite{KondoGaAsEcho,OnoGaAsEcho,IshiharaGaAsEcho}) and $^{71}$Ga
(non-selective echo on QWs \cite{Sanada} and QDs
\cite{MakhoninNatMat}) is shown in Fig. \ref{fig:CTEcho}(b) by the
triangles. It can be seen that the echo decay times in
self-assembled QDs are a factor of $\sim5-7$ larger compared to
lattice matched structures. Such increase in $T_2$ is due to
suppression of nuclear spin flip-flops and provides direct
evidence for the slow down of nuclear spin bath fluctuations in
the presence of spatially inhomogeneous QI.

In order to quantify the effect of QI on the nuclear spin bath
dynamics we turn to more detailed analysis of our experimental
results. At sufficiently large magnetic field along $Oz$ (above a
few mT) the interaction between any two nuclear spins $I$ and $J$
is described by the ''truncated'' dipole-dipole Hamiltonian
\cite{AbrahamBook}:
\begin{eqnarray}
\hat{H}_{dd}=\nu_{dd}\left[\hat{I}_{z}\hat{J}_{z}-\frac{1}{2}(\hat{I}_{x}\hat{J}_{x}+\hat{I}_{y}\hat{J}_{y})\right],\label{eq:Hdd}
\end{eqnarray}
where $\hat{I}$ and $\hat{J}$ are the spin operators, and the
coupling strength $\nu_{dd}$ depends on nuclei type and mutual
position ($\nu_{dd}\lesssim200$~Hz in frequency units for nearest
neighbors in InGaAs, and scales as $\propto r^{-3}$ with
internuclear distance $r$). The
$(\hat{I}_{x}\hat{J}_{x}+\hat{I}_{y}\hat{J}_{y})$ term enables
spin exchange flip-flops between nuclei $I$ and $J$: $(I_z,
J_z)\leftrightarrow(I_z\pm1, J_z\mp1)$, the process ultimately
responsible for electron spin decoherence via spectral diffusion.
A flip-flop can only happen if $I$ and $J$ have similar Zeeman
energies requiring them to be of the same isotope. If, however,
these two nuclei are subject to significantly different
quadrupolar shifts $\nu_{Q,I}$ and $\nu_{Q,J}$, so that
$|\nu_{Q,I}-\nu_{Q,J}|\gg\nu_{dd}$ the flip-flops will become
energetically forbidden, resulting in a slow down of nuclear spin
bath dynamics and potential increase in electron qubit coherence
time \cite{KorenevQ}.

Despite the very simple structure of the Hamiltonian of Eq.
\ref{eq:Hdd}, the calculation of the nuclear spin bath dynamics in
a crystal is a very difficult task due to the many-body nature of
the problem (each nuclear spin interacts with all other spins).
When arbitrary inhomogeneous QI is added the problem becomes
unsolvable in practice. However, for the limiting cases of very
small and very large QI the nuclear spin echo decay times can be
calculated with $\sim25\%$ accuracy from the first principles
using the method of moments \cite{Haase1993Echo}. The details of
the calculation techniques are discussed in the Methods and
further in Supplementary Sec. S4; in what follows we present the
results of these calculations and use them to analyze the
experimental data.

When quadrupolar shifts are much smaller than the dipolar
interaction $\nu_Q^{(2)}\ll\nu_Q^{(1)}\ll\nu_{dd}$, the nuclear
flip-flops are not affected by QI. The $T_2$ times calculated for
that case are shown in Fig. \ref{fig:CTEcho} with dashed lines for
different isotopes. These calculated values are in good agreement
with experiments on lattice matched GaAs/AlGaAs structures,
confirming the validity of the model employed.

In the opposite case of very strong inhomogeneous QI
$\nu_Q^{(1)}\gg\nu_Q^{(2)}\gg\nu_{dd}$ the dipole-dipole
flip-flops become energetically forbidden. This effect can be
described by truncating the flip-flop term in Eq. \ref{eq:Hdd},
leaving only the diagonal $\hat{I}_{z}\hat{J}_{z}$ term in the
Hamiltonian. The $\hat{I}_{z}\hat{J}_{z}$ term conserves the
nuclear polarization along the $Oz$ axis. But for polarization
orthogonal to $Oz$ (as is in a spin-echo experiment after the
initial $\pi/2$ pulse) this term causes decoherence
\cite{zzEchoDecay}, resulting in nuclear spin echo decay
\cite{Haase1993Echo} on a timescale which we denote as $T_{2,zz}$.
The $T_{2,zz}$ times calculated for the studied InGaAs dots are
shown by the solid lines in Fig. \ref{fig:CTEcho}. The $T_{2,zz}$
sets an upper limit on the echo decay time $T_{2}$.

We now turn to the question of how nuclear spin $T_2$ measurements
can be used to predict the effect of the nuclear spin bath
fluctuations on electron spin coherence. If the nuclear flip-flops
were completely suppressed the electron spin would experience only
a static nuclear field which cannot cause irreversible electron
spin decoherence -- a very attractive scenario for applications of
quantum dots in quantum information processing. But as explained
above the nuclear spin-echo $T_2$ time in that case does not tend
to infinity, it rather tends to a finite maximum value given by
$T_{2,zz}$. Thus in order to establish how strong the flip-flop
suppression is, we examine how close the experimental $T_2$ value
is to the calculated $T_{2,zz}$. For that we introduce a
characteristic nuclear spin flip-flop time $T_{2,ff}$ defined as
\begin{eqnarray}
T_{2,ff}^{-1}=T_{2}^{-1}-T_{2,zz}^{-1},\label{eq:T2ff}
\end{eqnarray}
so that in a quantum dot with $T_{2,ff}\rightarrow\infty$ we would
expect electron spin coherence not limited by nuclear spins.

It can be seen in Fig. \ref{fig:CTEcho} that $T_2<T_{2,zz}$ for In
and both Ga isotopes implying only a partial suppression of the
nuclear spin flip-flops. Using Eq. \ref{eq:T2ff} we calculate
$T_{2,ff}\sim5$~ms for all three of those isotopes at low fields
($B_z<2$~T). This $T_{2,ff}$ is $\sim$3 - 8 times larger than it
would have been in InGaAs/GaAs structures without strain
($T_{2,ff}$ values for this case are calculated to be
$T_{2,ff}\sim0.6$~ms, 1.5~ms, 1.1~ms for $^{115}$In, $^{69}$Ga and
$^{71}$Ga, see details in Supplementary Section S4). We also note
that the $T_2$ of $^{71}$Ga and $^{115}$In decreases with
increasing $B_z$ in agreement with the fact that second order
quadrupolar shifts depend on magnetic field as
$\nu_Q^{(2)}\propto1/B_z$ (Ref. \cite{AbrahamBook}), so that large
$B_z$ re-enables the nuclear flip-flops between the $I_z=\pm1/2$
spin states.

A very different picture is observed in Fig. \ref{fig:CTEcho} for
arsenic nuclei: the values of $T_{2}$ measured at $B_z=2-8$~T
coincide with the calculated $T_{2,zz}$ within the experimental
error. Eq. \ref{eq:T2ff} in that case diverges and gives
infinitely large $T_{2,ff}$ values, but we can conclude that
$T_{2,ff}\gg5$~ms for arsenic, implying very strong flip-flop
suppression. These findings are consistent with the spectroscopic
data in Fig. \ref{fig:NMRSpec} (b), where the CT spectra of
$^{75}$As are found to be $\sim$10 times broader than for gallium
nuclei, revealing much larger second order quadrupolar shifts of
arsenic nuclei, which are responsible for the strong suppression
of the nuclear spin exchange flip-flops.

The second order shifts $\nu_Q^{(2)}$ appear whenever $\nabla
\vec{E}$ is not a cylindrically-symmetric tensor with its main
axis along $B_z$ (Ref. \cite{AbrahamBook}). One obvious reason for
low-symmetry $\nabla \vec{E}$ in self-assembled QDs, is
non-uniaxial symmetry of the elastic strain tensor, or deviation
of the strain main axis from $B_z$. Such a mechanism is likely to
be the main cause of the CT inhomogeneous broadening of gallium
and indium, resulting in the above increase in $T_{2,ff}$ by a
factor of $\sim$3 - 8.

For the anion $^{75}$As additional $\nu_Q^{(2)}$ shifts are
induced by random alloy mixing of the cationic Ga and In atoms:
Each arsenic nucleus has four nearest neighbors, and unless all of
them are of the same type (all gallium or all indium) a non-zero
$\nabla \vec{E}$ will appear \cite{KnijnIntermixing}. Furthermore,
unlike the elastic strain fields that change gradually over many
crystal unit cells, the configuration of the neighboring atoms is
random, so that even the nearest arsenic nuclei (which have the
strongest dipolar coupling) can have very different $\nu_Q^{(2)}$.
Such compositional disorder can induce large spatially
inhomogeneous CT frequency shifts, drastically suppressing the
flip-flops. We propose that such effects can be used to engineer
QDs with a frozen nuclear spin bath. One possible approach is to
substitute some of the arsenic nuclei with antimony and/or
phosphorus: in such InGaAsSb(P) quantum dots gallium and indium
spins will also experience large inhomogeneous $\nu_Q^{(2)}$
shifts due to atomic-scale alloy disorder, resulting in an overall
slow down of the flip-flops for all isotopes.

The increase of the nuclear spin-echo $T_2$ observed for the
central transitions is driven by the second order quadrupolar
shifts $\nu_Q^{(2)}$ which are significantly smaller than the
first order shifts $\nu_Q^{(1)}$ ($\nu_Q^{(2)}$$\sim$10 - 100~kHz
compared to $\nu_Q^{(1)}$$\sim$1 - 10~MHz). Therefore, we expect
that the flip-flops of the nuclei in $|I_z|>1/2$ states (affected
by $\nu_Q^{(1)}$) are effectively frozen for all isotopes in
strained InGaAs dots. Consequently, electron spin decoherence in
self-assembled dots is caused solely by the flip-flops of the
nuclei in $I_z=\pm1/2$ states.

Finding an exact relation between the nuclear spin bath coherence
times and the central spin coherence times is a complicated
problem, in particular when quadrupolar interactions are involved
\cite{Sinitsyn2012}. However, some general qualitative conclusions
can be readily drawn. Firstly, it has been shown that for quantum
dots the spectral diffusion is well into the ''slow bath'' regime
(as opposed to ''motional narrowing'') \cite{DasSarma2003}, thus
the extended nuclear spin coherence times reported here are
expected to result in longer central spin coherence times.
Secondly, dynamic nuclear spin polarization \cite{GammonPRL}, that
enhances the occupancy of the $|I_z|\approx I$ states can be used
to effectively depopulate and ''dilute'' the $I_z=\pm1/2$ states
resulting in further reduction of the nuclear spin fluctuations.
Based on recent observation of electron spin coherence times
$>200$~$\mu$s in lattice matched GaAs/AlGaAs QDs \cite{Bluhm}, we
may anticipate even longer, millisecond-range coherence times for
self-assembled InGaAs/GaAs dots.

In conclusion we have demonstrated the first direct probing of the
coherent nuclear spin bath dynamics in strained quantum dots. We
predict that electron spin qubits in self-assembled structures
exhibiting large quadrupolar interactions have a significant
advantage over the lattice-matched counterparts. Engineering of
strain and alloy disorder can be used to enhance this advantage
further and open the way for optically-active quantum dot qubits
with completely frozen nuclear spin environment. Future progress
in this direction will depend strongly on advances in theoretical
modeling of the central spin coherence in presence of quadrupolar
effects \cite{Sinitsyn2012}, a problem which previously could not
be fully grasped due to the lack of experimental knowledge on the
nuclear spin coherence. The techniques for pulsed NMR in strained
quantum dots developed here are not restricted to spin-echo and
can be readily extended to accommodate the whole variety of pulse
sequences used in Fourier transform NMR, offering a powerful tool
to explore the many-body physics of interacting nuclear spins in
strained nanostructures.

\section*{\label{sec:Methods} Methods Summary}

\textbf{Continuous wave NMR spectroscopy.} The CT spectra of an
individual InGaAs/GaAs QD shown in Fig. \ref{fig:NMRSpec}(b) were
measured using ''inverse'' method which provides $>$8 times CT
signal enhancement for $I=3/2$ nuclei \cite{QNMRArxiv}. The NMR
signal is calculated as the hyperfine shift of the quantum dot
Zeeman doublet divided by the spectral ''gap'' width, so that the
values on the vertical scale give the spectral density of the
distribution of the nuclear resonance frequencies. The spectral
gap width (determining the spectral resolution) is 6~kHz for the
$^{69}$Ga spectra, and 16~kHz (32~kHz) for the $B_z=8$~T
($B_z=2$~T) spectrum of $^{75}$As. For convenience the spectra are
plotted as a function of $\nu-\nu_0$, where $\nu_0$ is a constant
proportional to the isotope gyromagnetic ratio:
$\nu_0/B_z\approx7.33$~MHz/T for $^{75}$As and $\nu_0/B_z\approx
10.3$~MHz/T for $^{69}$Ga.

\textbf{Techniques for pulsed NMR measurements.} We implement
optically detected pulsed NMR techniques which extend the
techniques and are based on the results of our previous work of
Ref. \cite{QNMRArxiv}. The timing diagram of one measurement cycle
and the changes to nuclear spin polarization are shown
schematically in Fig. \ref{fig:CTScheme}(a) (spin $I=3/2$ is used
as an example). The cycle starts with optical pumping using high
power $\sigma^+$ circularly polarized laser [stage (1)]. Spin
polarized electrons excited by the laser transfer their
polarization to nuclear spins via the hyperfine interaction
\cite{GammonPRL,HoleNucIso}. Pump duration is chosen long enough
($\sim$3 - 7~s depending on magnetic field $B_z$) to achieve the
steady state nuclear spin polarization degrees exceeding 50\%,
which means that a large portion of the nuclei is initialized into
the $I_z=-3/2$ state. In order to make the NMR signal of the
central transition detectable the population of the
$I_z=-1/2(+1/2)$ state must be maximized (minimized). This is done
at stage (2) using ''population transfer'' technique
\cite{Haase1993Enhance}: an rf field containing two frequency
components is applied, the frequencies are swept over both
satellite transition bands $-3/2 \leftrightarrow -1/2$ and $+1/2
\leftrightarrow +3/2$ resulting in adiabatic inversion of the
populations of the $-3/2$ and $-1/2$ states as well as $+1/2$ and
$+3/2$ states. Following that a sequence of rf pulses resonant
with the CT is applied [stage (3)]. Different sequences can be
implemented, depending on the experiment: a single pulse of a
variable duration is used for Rabi-oscillation measurements [Fig.
\ref{fig:CTScheme}(b)], while a three-pulse sequence is used to
measure either the spin-echo [Fig. \ref{fig:CTScheme}(c),
$\pi/2-\tau_0-\pi-\tau-\pi/2$ sequence with $\tau_0$ fixed to
0.4~ms] or spin-echo decay [Fig. \ref{fig:CTScheme}(d),
$\pi/2-\tau-\pi-\tau-\pi/2$ sequence]. The rf amplitude is chosen
to give $\pi/2$ phase rotation for 3 - 8~$\mu$s long pulses
(depending on isotope). This corresponds to pulse bandwidths of
$\sim$100~kHz, and, since satellite transitions are shifted by the
much bigger ($\sim$1 - 10~MHz) first order quadrupolar shifts,
this ensures selective excitation of the central transition.
Finally [stage (4)] we probe the effect of the NMR pulse sequence
by measuring the changes in the average nuclear spin polarization
$\langle I_z \rangle$ on the single quantum dot. This is achieved
by exciting the dot with a short ($\sim$1 - 4~ms depending on
$B_z$) probe laser pulse and measuring the hyperfine shifts in the
QD photoluminescence spectrum. In order to improve the signal to
noise ratio the experimental cycle is repeated 20 - 50 times for
each parameter value [e.g. for each value of $2\tau$ in Fig.
\ref{fig:CTScheme}(d)]. Further details of experimental techniques
can be found in Supplementary Sections S2.

\textbf{Theoretical model.} Nuclear spin decoherence is a result
of nuclear-nuclear spin interactions: each individual nuclear spin
has its own spin environment producing additional magnetic field,
which changes the resonant frequency of that nucleus. Thus the
problem of calculating the nuclear spin decoherence is equivalent
to the problem of calculating homogeneous NMR line broadening. In
principle this problem can be solved by diagonalizing the
Hamiltonian of the nuclear-nuclear interactions. This however is
practically impossible even for a system of few tens of spins, let
alone the whole crystal. An insightful solution to this difficulty
has been found by Van Vleck \cite{AbrahamBook,VanVleck} who showed
that the moments of the NMR lineshape can be expressed as traces
of certain quantum mechanical operators. The key property of the
trace is that it can be calculated in any wavefunction basis,
hence diagonalization of the Hamiltonian is not needed. This
technique does not allow an exact resonance lineshape to be found,
but in most cases the second moment $M_2$ (corresponding to the
homogeneous NMR linewidth) contains sufficient information.

The nuclear spin coherence time can be estimated as
$T_2\approx\sqrt{2/M_2}$. The calculation of $M_2$ for a whole
crystal is a straightforward but very tedious process involving
summation of various matrix elements. If one wants to calculate
the spin-echo coherence time (as opposed to the free-induction
decoherence time), some of the matrix elements must be discarded
from the summation. Quadrupolar interaction is also taken into
account by further ''truncation'' of the sums. The details of
these calculations can be found in Refs.
\cite{VanVleck,AbrahamBook,Haase1993Echo} and are also outlined in
Supplementary Sec. S4.

\textbf{ACKNOWLEDGMENTS} The authors are grateful to G. Burkard,
E. Welander, L. Cywinski and K. V. Kavokin for fruitful
discussion. This work has been supported by the EPSRC Programme
Grant EP/J007544/1 and the Royal Society. E.A.C. was supported by
a University of Sheffield Vice-Chancellor's Fellowship.

\textbf{ADDITIONAL INFORMATION} Correspondence and requests for
materials should be addressed to E.A.C.
(e.chekhovich@sheffield.ac.uk).


\renewcommand{\thesection}{S\arabic{section}}
\setcounter{section}{0}
\renewcommand{\thefigure}{S\arabic{figure}}
\setcounter{figure}{0}
\renewcommand{\theequation}{S\arabic{equation}}
\setcounter{equation}{0}
\renewcommand{\thetable}{S\arabic{table}}
\setcounter{table}{0}

\renewcommand{\citenumfont}[1]{S#1}
\makeatletter
\renewcommand{\@biblabel}[1]{S#1.}
\makeatother

\pagebreak \pagenumbering{arabic}

\section*{Supplementary Information}

The document consists of the following sections:\\
\ref{SI:Samples}. Quantum dot sample structure\\
\ref{SI:Techniques}. Details of pulsed NMR techniques\\
\hspace*{0.5cm}\ref{SI:TechniquesTDiagr}. Pump-probe techniques for optically detected pulsed NMR\\
\hspace*{0.5cm}\ref{SI:TechniquesHardware}. Pulsed optically detected NMR: the hardware\\
\ref{SI:Experiment}. Additional experimental results\\
\ref{SI:CalcT2}. First principle calculation of the nuclear spin-echo decay time $T_2$\\
\hspace*{0.5cm}\ref{SI:CalcT2i}. Nuclear spin echo decay under strongly inhomogeneous quadrupolar shifts [case (i)]\\
\hspace*{0.5cm}\ref{SI:CalcT2ii}. Nuclear spin echo decay under homogeneous quadrupolar shifts [case (ii)]\\

\section{\label{SI:Samples}Quantum dot sample structure}

The InGaAs/GaAs sample has been described previously in Refs.
\cite{InGaAsSamp2S,InGaAsSamp3S,LowPowDNPS} and in The
Supplementary Information of Refs. \cite{QNMRArxivS,HoleNucIsoS}.
The sample consists of a single layer of nominally InAs quantum
dots (QDs) placed within a microcavity structure which is used to
select and enhance the photoluminescence from part of the
inhomogeneous distribution of QD energies. The sample was grown by
molecular beam epitaxy. The QDs were formed by deposition of
~1.85~monolayers (MLs) of InAs - just above that required for the
nucleation of dots. As a result, we obtain a low density of QDs at
the post-nucleation stage. The cavity Q factor is $\sim$250 and
the cavity has a low temperature resonant wavelength at around
920~nm.

\section{\label{SI:Techniques}Details of pulsed NMR techniques}

We implement optically detected pulsed NMR techniques which extend
the techniques of our previous work \cite{QNMRArxivS,HoleNucIsoS}.
Some description of experimental techniques was previously
reported in the Supplementary of Ref. \cite{QNMRArxivS} and also
applies to this work. Below we focus on the techniques specific to
the present work.

\subsection{\label{SI:TechniquesTDiagr}Pump-probe techniques for optically detected pulsed NMR}

\begin{figure}
\includegraphics[bb=74pt 96pt 497pt 183pt]{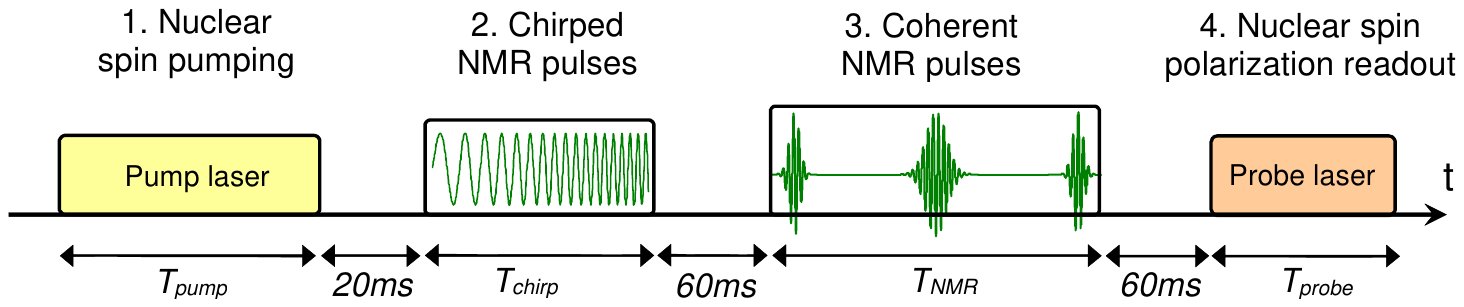}
\caption{\label{fig:TDiagrS} \textbf{Timing diagram for pulsed NMR
experiments on strained quantum dots.} See detailed explanation in
Sec. \ref{SI:TechniquesTDiagr}.}
\end{figure}

The timing diagram of one measurement cycle is shown schematically
in Fig. \ref{fig:TDiagrS} (this is a detailed version of diagram
in Fig. 2(a) of the main text). The cycle consists of the
following four stages:

\textit{Stage 1.} In order to achieve sufficiently large NMR
signal the nuclei must be prepared in a highly-polarized state.
Manipulation of nuclear spin polarization relies on the hyperfine
interaction of electrons and nuclear spins. Excitation with a
$\sigma^+$ circularly polarized ''pump'' lazer generates spin
polarized electrons, which transfer their polarization to nuclear
spins via the hyperfine interaction
\cite{GammonPRLS,TartakovskiiS,LaiS,EbleS,Braun1S,InPX0S}. In this
work dynamic nuclear spin polarization (polarization degrees
exceeding 50\%) is achieved using high power nonresonant optical
pumping \cite{LowPowDNPS}. The pumping wavelength of $\sim$850~nm
corresponds to excitation into the QD wetting layer states.
Optical powers exceeding the QD saturation level by more than a
factor of 10 are typically used \cite{InPX0S,LowPowDNPS}. At these
powers the dependence of the steady-state nuclear polarization on
the optical power saturates. This ensures large nuclear spin
polarization degree as well as its good reproducibility due to
insensitivity to laser power fluctuations. Pump durations of
$T_{pump}=3-7$~s (depending on magnetic field $B_z$) are used to
achieve a steady-state polarization independent of the
polarization left after the previous cycle. A delay of 20~ms is
introduced after the pumping to ensure that the pump laser is
completely blocked by a mechanical shutter, so that NMR on a
quantum dot is measured in the dark.

\textit{Stage 2.} The amplitude of the NMR signal of the central
transition (CT) studied in this work is determined by the
difference in population probabilities of the $I_z=-1/2$ and
$I_z=+1/2$ nuclear spin states. However, for spin $I>1/2$ inducing
large nuclear spin polarization degree does not necessarily
increase the CT signal. In fact, in the limit of 100\%
polarization all nuclei will be in the $I_z=-I$ state and the NMR
signal of CT will vanish. Simple analysis shows that any optically
induced (Boltzmann) distribution would yield CT NMR signal too
weak to detect in our setup. Thus nuclear spin level populations
have to be manipulated artificially. We achieve this via
''population transfer'' techniques \cite{Haase1993EnhanceS} which
uses chirped radio-frequency (rf) pulses. For spin-3/2 nuclei we
use an rf field containing two spectral components. The
frequencies of both components are swept ''outwards'' from
$\nu_{CT}\pm\nu_i$ to $\nu_{CT}\pm\nu_f$, where $\nu_{CT}$ is the
frequency of the central transition. The initial frequency offset
is chosen to be $\nu_i=20$~kHz, whereas the final offset is chosen
to be larger than the maximum first-order quadrupolar shift of the
studied isotope ($\nu_f=8.5$~MHz for $^{75}$As, $\nu_f=5.6$~MHz
for $^{69}$Ga and $\nu_f=3.5$~MHz for $^{71}$Ga are used). Such
frequency sweeps adiabatically swap the populations of the $-3/2$
and $-1/2$ states as well as $+1/2$ and $+3/2$ states,
significantly enhancing the population difference of the $\pm1/2$
states. For spin-9/2 indium the sweeps are done ''inwards''
($\nu_i>\nu_f$), with $\nu_i=8.5$~MHz and $\nu_f=120-230$~kHz,
which effectively transfers the populations of the $\pm9/2$ states
to the $\pm1/2$ states. The amplitude of each swept component is
$\sim0.2-0.5$~mT. The sweep rate is calibrated in an additional
measurement and is chosen to maximize the CT signal. Typical rates
used are $\sim7-20$~MHz/s, so that the duration of the chirped
pulse is $T_{chirp}\sim0.3-1.5$~s. After the sweep a 60~ms delay
is introduced to permit any transverse nuclear magnetization to
decay and allow a mechanical relay to switch between the ''chirp''
and ''pulse'' rf signal sources (see Section
\ref{SI:TechniquesHardware}).

\textit{Stage 3.} A sequence of rf pulses resonant with the
central transition is applied to manipulate coherently the
magnetization of the $I_z=\pm1/2$ nuclear spin subspace. Different
sequences can be implemented, e.g. Rabi-oscillations, Hahn-echo or
echo decay, as demonstrated in Fig. 2 of the main text. The rf
amplitude is chosen to give $\pi/2$ phase rotation of the
$I_z=\pm1/2$ subspace for 3-8~$\mu$s long pulse (depending on
isotope). All pulse sequences are designed in a way that the final
magnetization that needs to be measured is projected on to the
$Oz$ axis, so that it can be detected optically. For example in
the echo decay sequence $\pi/2-\tau-\pi-\tau-\pi/2$ the last
$\pi/2$ pulse rotates the transverse magnetization (which is of
interest) aligning it along the $Oz$ axis. The total duration of
the NMR pulse sequence $T_{NMR}$ varies from a few microseconds to
110~ms. After the pulse sequence a 60~ms delay is introduced to
allow the decay of any spurious transverse nuclear magnetization
and the dissipation of heating induced by chirped or resonant NMR
pulses.

\textit{Stage 4.} Finally we probe the effect of the NMR pulse
sequence by measuring the changes in the average nuclear spin
polarization $\langle I_z \rangle$ on the dot. This is achieved by
exciting the dot with a short ($T_{probe}=1-4$~ms depending on
$B_z$) nonresonant ($\sim850$~nm) probe laser pulse and measuring
the hyperfine shifts of the Zeeman splitting in the QD
photoluminescence spectrum
\cite{GammonPRLS,QNMRArxivS,HoleNucIsoS}. The power of the probe
laser is $\sim1/10$ of the QD saturation power. We use
differential measurements: for spin-echo and echo decay
experiments the NMR signal is calculated as the difference of the
QD Zeeman splitting measured with the $\pi/2-\tau-\pi-\tau-\pi/2$
sequence and the splitting measured with the $\pi/2-100$~ms$-\pi$
sequence. In this way a complete decay of echo corresponds to
$\sim0~\mu$eV signal. For Rabi-oscillations measurements we
subtract the Zeeman splitting obtained from a measurement with a
single $\pi/2$ pulse.

The total duration of stages $2-4$ does not exceed $\sim2$~s,
which is much shorter than the nuclear spin polarization
longitudinal ($T_1$) decay time ($>$1 hour). In order to improve
the signal to noise ratio the experimental cycle is repeated
$10-50$ times during the photoluminescence spectrum acquisition
for each parameter value [e.g. for each value of $2\tau$ in the
spin echo measurements].

\subsection{\label{SI:TechniquesHardware}Pulsed optically detected NMR: the hardware}

\begin{figure}
\includegraphics[bb=92pt 59pt 377pt 386pt]{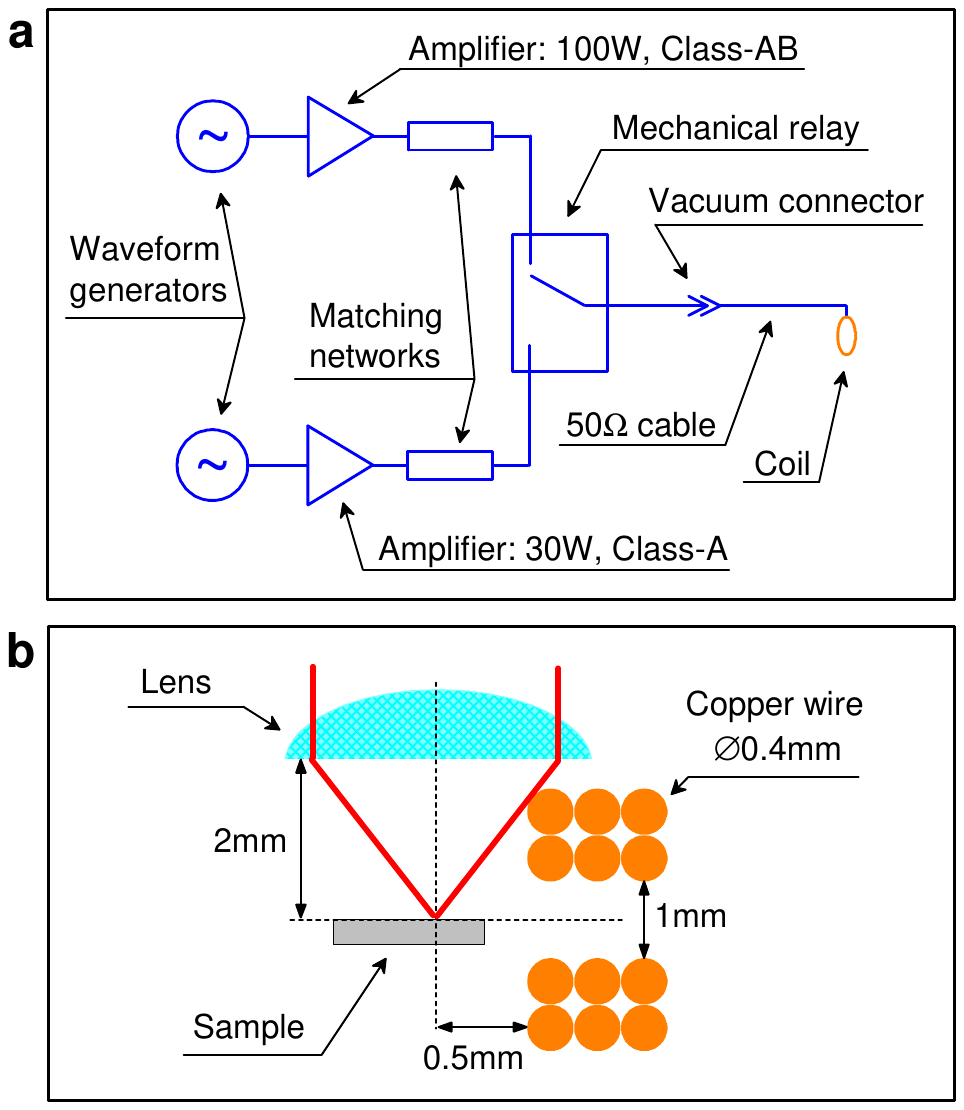}
\caption{\label{fig:Hardware} \textbf{Hardware implementation of
the pulsed NMR techniques.} \textbf{a,} Circuit diagram.
\textbf{b,} Schematic drawing of the NMR coil arrangement (side
view). See detailed explanation in Sec.
\ref{SI:TechniquesHardware}.}
\end{figure}

A schematic circuit diagram of the NMR setup is shown in Fig.
\ref{fig:Hardware}(a). The ''chirp'' and the ''resonant'' pulses
are generated in separate arms, each containing a digital
arbitrary/function generator, a high power amplifier and a
matching network. The signal from one of the two arms is selected
by a mechanical relay and is then transmitted into the low
temperature vacuum insert via a coaxial cable. All equipment and
cables have 50~$\Omega$ impedance, while the coil impedance
differs significantly from 50~$\Omega$, hence the need for the
matching networks. For resonant NMR pulses only a narrow bandwidth
is required (few hundred kHz); thus we use a single stub network
which gives nearly ideal impedance matching at a specific
frequency. The ''chirped'' pulse requires a much larger bandwidth
(up to 20~MHz), over which it is impossible to achieve good
impedance matching. In that case the role of the matching network
(consisting of lumped $LC$ elements and cables of different
length) is to provide nearly constant transmission over the
frequency sweep band. The large mismatch of such a network is
compensated by the use of a mismatch-tolerant class-A power
amplifier.

A schematic drawing of the NMR coil arrangement is shown in Fig.
\ref{fig:Hardware}(b). The coil consists of 6 turns of a copper
wire (diameter 0.4~mm) wound in two layers. The external radius of
the coil is smaller than the working distance of the lens used to
excite and collect photoluminescence. As a result the coil can be
positioned very close to the edge of the quantum dot sample: the
distance between the coil edge and the lens focal point is
$\sim$0.5~mm. The lens and the coil are fixed while the sample is
mounted on an XYZ piezo-positioner allowing the sample surface to
be scanned and different individual dots to be studied. For the
range of frequencies used (5-110 MHz) the coil produces an
oscillating magnetic field $\sim$20~mT for an input power of
100~W.

\section{\label{SI:Experiment}Additional experimental results}

\begin{figure}
\includegraphics[bb=38pt 78pt 409pt 289pt]{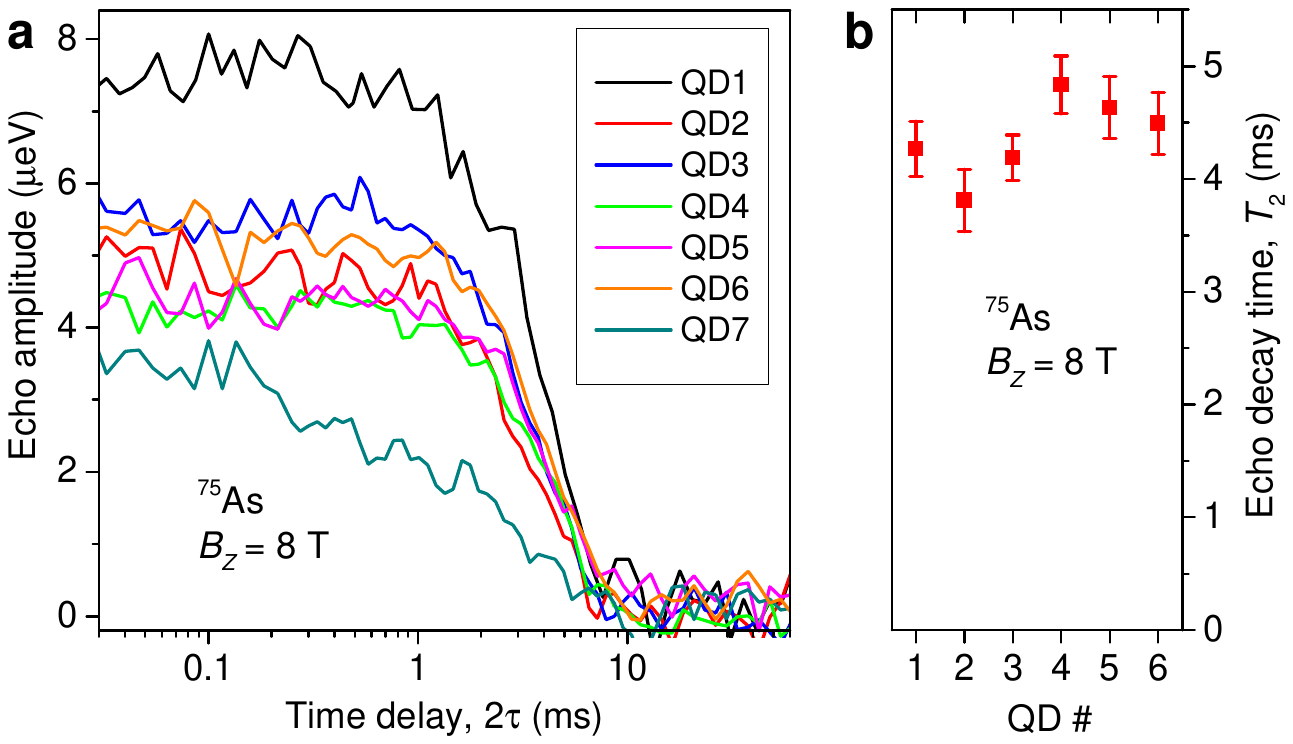}
\caption{\label{fig:EchoStat} \textbf{Additional echo decay
measurements.} \textbf{a,} Spin echo decay curves of $^{75}$As
measured for dots QD1-QD7 at $B_z=8$~T. \textbf{b,} Spin echo
decay times $T_2$ obtained for dots QD1-QD6 using Gaussian fitting
of the curves in \textbf{a} (90\% confidence intervals).}
\end{figure}

All experimental results presented in the main text were measured
on the same quantum dot QD1. To verify the consistency of our
conclusions we have carried out spin echo decay measurements on an
additional set of different dots (QD2-QD7) from the same sample.
Spin echo decay curves of $^{75}$As measured for dots QD1-QD7 at
$B_z=8$~T are shown in Fig. \ref{fig:EchoStat}(a) with different
colours.

Quantum dots QD1-QD6 demonstrate echo decay that can be well
described by a Gaussian function. By contrast, one of the dots
(QD7) demonstrates significantly faster non-Gaussian decay.

The echo decay times $T_2$ obtained from Gaussian fitting are
shown in Fig. \ref{fig:EchoStat}(b) for QD1-QD6. There is some
dot-to-dot variation. However, it is comparable to the
experimental error, and all of the experimental $T_2$ values are
also in good agreement with the calculated decay time of
$\sim$4.03~ms, confirming the reproducibility of the presented
experimental results and analysis.

The considerably faster echo decay observed for QD7 is not fully
understood. One possibility is that additional nuclear spin
decoherence is caused by charge fluctuations. Such fluctuations
can not be controlled in our experemints, since we are using
electric-gate-free structures. Thus electrons or holes can hop
between the nearby impurities or can randomly populate the dot.
This might also be the cause for some dot-to-dot variations in
$T_2$ values and will be a subject of further studies (e.g. using
dots in Schottky-diode structures).

\section{\label{SI:CalcT2}First principle calculation of the nuclear spin-echo decay time $T_2$}

Our calculations are done within a framework developed by Haase
and Oldfield in Ref. \cite{Haase1993EchoS}. Their model is based
on the well-known method of moments initially developed by Van
Vleck \cite{VanVleckS}. Below we outline the key points of this
model and show how it is used to derive the results presented in
the main text.

The decay of the nuclear spin echo is caused by the dipole-dipole
interaction between nuclear spins. At sufficiently large magnetic
field along $Oz$ (above a few mT) the interaction between two
nuclear spins $I$ and $J$ is described by the ''truncated''
dipole-dipole Hamiltonian\cite{AbrahamBookS}:
\begin{align}
&\hat{H}_{dd}/(2\pi\hbar)=\nu_{dd}\left[\hat{I}_{z}\hat{J}_{z}-\frac{1}{2}(\hat{I}_{x}\hat{J}_{x}+\hat{I}_{y}\hat{J}_{y})\right],\nonumber\\
&\nu_{dd}=\frac{\mu_0}{4\pi}\frac{\hbar}{2\pi}\gamma_I\gamma_J\frac{1-3\cos^2\theta}{r^3},\label{eq:Hdd}
\end{align}
where $\hat{I}$ and $\hat{J}$ are the nuclear spin operators, and
$\nu_{dd}$ is the coupling strength. $\nu_{dd}$ depends on the
gyromagnetic ratio $\gamma_I$ ($\gamma_J$) of the spin $I$($J$),
the internuclear distance $r$ and the angle $\theta$ between the
magnetic field and the vector $\vec{r}$ connecting the nuclei
($\nu_{dd}\lesssim200$~Hz in frequency units for all nuclei in
InGaAs sample).

The interactions between nuclei of the same type (homonuclear
coupling) and the nuclei of different types (heteronuclear
coupling) have a different role in nuclear spin echo decay. The
dephasing caused by the homonuclear coupling is not refocused by
the $\pi$ pulse and, as a result, the spin echo decays. The
heteronuclear coupling between the ''studied'' spins of isotope
$I$ and the spins of ''another'' isotope $S$, can be viewed as a
randomly distributed magnetic field induced by isotope $S$ and
acting on $I$. On sufficiently short time-scales this effective
field is static and thus its effect is refocused by the $\pi$
pulse. However, the spin flip-flops induced by the homonuclear
coupling of the $S$ nuclei will make this effective magnetic field
time-dependent. As a result the spin echo of isotope $I$ will
decay via the spectral diffusion mechanism (this is exactly the
same mechanism that causes the decoherence of the electron spin
interacting with the nuclear spin bath
\cite{DasSarma2003S,DasSarma2006S,Sham2006S}). Below we will
consider both homonuclear and heteronuclear interaction.

Quadrupolar interactions modify nuclear spin spectra and affect
nuclear spin dephasing dynamics (for introduction to quadrupolar
effects see Chapter X of Ref. \cite{SlichterBookS}). An exact
treatment of nuclear spin echo decay under the combined effect of
quadrupolar and dipole-dipole Hamiltonians is unattainable.
However, the echo decay times can be obtained relatively easily
for two limiting cases\cite{Haase1993EchoS}: (i) The quadrupolar
interaction is strongly inhomogeneous, i.e. the difference in
quadrupolar shifts (including the second-order shifts) of the
nearby nuclei is much larger than their dipolar coupling. Under
these conditions all dipolar nuclear flip-flops are energetically
forbidden. Such a case would correspond to a self-assembled
quantum dot where first-order quadrupolar shifts are strongly
inhomogeneous due to inhomogeneous elastic strain, while
second-order shifts are inhomogeneous due to atomic-scale
disorder. (ii) The opposite case is observed for homogeneous
quadrupolar interactions, i.e. each nucleus of a particular
isotope is subject to the same electric field gradient. In that
case the flip-flops between nuclear spins $I$ and $J$ of the type
$(I_z=m,J_z=m\pm1)\leftrightarrow(I_z=m\pm1,J_z=m)$ are
energetically allowed. This case is relevant for uniformly
strained GaAs/AlGaAs quantum
wells\cite{KondoGaAsEchoS,OnoGaAsEchoS,IshiharaGaAsEchoS}. We now
consider these two cases in detail.

\subsection{\label{SI:CalcT2i}Nuclear spin echo decay under strongly inhomogeneous quadrupolar shifts [case (i)].}

This case is the easiest for numerical calculations. Since all
flip-flops are forbidden, the only source of nuclear spin echo
decay is due to the diagonal ($\hat{I}_{z}\hat{J}_{z}$) part of
the homonuclear dipole-dipole interaction (heteronuclear coupling
has no effect since there are no flip-flops that can cause
spectral diffusion).

The nuclear spin echo decay time $T_2$ (the time during which the
echo amplitude decays to $1/e$ of its initial value) due to the
homonuclear coupling can be approximated as\cite{Haase1993EchoS}:
\begin{equation}
T_2=\sqrt{2/M_{2E}}\label{eq:T2M2},
\end{equation}
where $M_{2E}$ is the second moment of the dipole-dipole
interaction (Hamiltonian Eq. \ref{eq:Hdd}) calculated for the
whole crystal:
\begin{equation}
M_{2E}=F\left(\frac{\mu_0}{4\pi}\right)^2\hbar^2\gamma^4a^{-6}\rho\sum_{i\ne0}\left(\frac{3}{2}\times\frac{1-3\cos^2\theta_i}{(r_i/a)^3}\right)^2\label{eq:M2E},
\end{equation}
where $\mu_0$ is vacuum magnetic permeability (introduced here to
carry out calculations in SI units), $\hbar$ - Planck's constant,
$\gamma$ - nuclear gyromagnetic ratio. The summation goes over the
whole crystal, $r_i$ is the length of the vector $\vec{r_i}$
between the $i$-th nucleus and the nucleus chosen as an origin
($i$=0), $\theta_i$ is the angle between $\vec{r_i}$ and the
direction of magnetic field, $a$ - is the lattice constant. $\rho$
- is the isotope abundance. Here we take into account that all
isotopes are sufficiently abundant ($\rho>0.1$) so that
$M_{2E}\propto\rho$ (Chapter IV, \S 9 of Ref.
\cite{AbrahamBookS}). The factor $F$ depends on the nature of the
quadrupolar shifts and the studied transition. For spin-echo decay
of the central transition under selective excitation and under
strongly inhomogeneous quadrupolar shifts [case (i)] it is:
\begin{equation}
F=\frac{2}{9(2 I+1)}\label{eq:Fzz},
\end{equation}

The fcc lattice sum is calculated numerically
\begin{eqnarray}
\sum_{i\ne0}\left(\frac{3}{2}\times\frac{1-3\cos^2\theta_i}{(r_i/a)^3}\right)^2\approx155.13\label{eq:fSumSame},
\end{eqnarray}

Since $^{75}$As is the only isotope of arsenic, we have $\rho=1$.
The relative concentration of indium and gallium in the studied
dots is estimated to be 0.24 and 0.76 respectively
\cite{QNMRArxivS}. Taking also into account the natural abundances
we have $\rho\approx0.46$ for $^{69}$Ga, $\rho\approx0.30$ for
$^{71}$Ga and $\rho\approx0.23$ for $^{115}$In. These and other
nuclear parameters are listed in Table \ref{tab:T2Calc}.

The low temperature lattice constant of GaAs is $a_0=0.564786$~nm.
The lattice constant of bulk InAs is larger by 7.17\%. To take
into account that only 24\% of atoms are indium we use linear
interpolation of lattice constant dependence on indium
concentration. Furthermore the lattice constant is modified by the
elastic strain. To estimate this effect we first need to imagine
that an InGaAs dot is uniformly compressed to match the lattice
constant of GaAs and is embedded into the unstrained GaAs matrix
in a defect-free manner \cite{DaviesQDStrainS}. Such structure is
not in equilibrium - the strain fields have to relax resulting in
quantum dot expansion. Theoretical analysis shows that a strained
InGaAs dot recovers $\sim2/3$ of the changes in size that were
initially induced by compression \cite{DaviesQDStrainS}. Thus for
the lattice constant we use the following estimate $a\approx
a_0(1+\frac{2}{3}\times0.24\times0.0717)\approx0.571$~nm.

Using these values we calculate the following echo decay times for
case (i) which we denote as $T_{2,zz}$: 4.03~ms for $^{75}$As,
3.04~ms for $^{69}$Ga, 2.31~ms for $^{71}$Ga and 8.08~ms for
$^{115}$In. These $T_{2,zz}$ values are also given in Table
\ref{tab:T2Calc} and are shown by the solid lines in Fig. 3 of the
main text.

\begin{table}
\caption{\label{tab:T2Calc} Nuclear spin parameters and calculated
nuclear spin echo decay times $T_2$}
\begin{ruledtabular}
\begin{tabular}{lccccc}  
Parameter & $^{75}$As  & $^{115}$In  & $^{69}$Ga & $^{71}$Ga \\
\hline
Nuclear spin $I$ & 3/2  & 9/2  & 3/2 & 3/2 \\
\hline
Gyromagnetic ratio $\gamma$ ($10^7$ s$^{-1}$/T)& 4.596  & 5.897  & 6.439 & 8.181 \\
\hline
Natural abundance  & 1  & 0.957 & 0.601 & 0.399 \\
\hline
Abundance in studied quantum dots $\rho$ & 1  & 0.23 & 0.46 & 0.3 \\
\hline\hline
Calculated $T_2$ times (ms) for the studied dots:\\
\hline
Strongly inhomogeneous quadrupolar shifts. $T_{2,zz}$  & 4.03 &  8.08 & 3.04 & 2.31 \\
\hline
Homogeneous quadrupolar shifts. $T_{2,zz+ff}$  & 1.3 &  0.55 & 0.98 & 0.75 \\
\hline
Homogeneous quadrupolar shifts, including\\
the effect of heteronuclear couplings. $T_{2,zz+ff+IS}$  & 0.93 &  0.42 & 0.66 & 0.51 \\
\end{tabular}
\end{ruledtabular}
\end{table}

\subsection{\label{SI:CalcT2ii}Nuclear spin echo decay under homogeneous quadrupolar shifts [case (ii)].}

Nuclear spin echo decay in case (ii) is caused by both homonuclear
and heteronuclear dipole-dipole couplings. First we consider the
homonuclear interaction: now in addition to the diagonal part
($\propto\hat{I}_{z}\hat{J}_{z}$), the flip-flop part
($\propto\hat{I}_{x}\hat{J}_{x}+\hat{I}_{y}\hat{J}_{y}$) of the
Hamiltonian eq. \ref{eq:Hdd} also contributes to echo decay. This
is taken into account by using a different value of $F$ in eq.
\ref{eq:M2E} which now reads \cite{Haase1993EchoS}:
\begin{equation}
F=\frac{2}{9(2I+1)}\times\frac{80I^4+160I^3+344I^2+264I+333}{256}\label{eq:Fzzff}.
\end{equation}
The resulting echo decay times (which we denote as $T_{2,zz+ff}$)
are 1.3~ms for $^{75}$As, 0.98~ms for $^{69}$Ga, 0.75~ms for
$^{71}$Ga and 0.55~ms for $^{115}$In (see also Table
\ref{tab:T2Calc}).

It follows from Eqns. \ref{eq:T2M2}, \ref{eq:Fzz}, \ref{eq:Fzzff}
that the ratio of homonuclear decay times in case (i) and case
(ii) depends only on spin $I$: $T_{2,zz}/T_{2,zz+ff}\approx3.1$
for $I=3/2$ and $T_{2,zz}/T_{2,zz+ff}\approx14.8$ for $I=9/2$. We
also note here that the factor $F$ determining the echo decay of
the central transition is most influenced by the inhomogeneity of
the second-order quadrupolar shifts. By contrast the nature of the
first-order quadrupolar shifts makes little difference: values of
$F$ differing by less than 10\% from that of Eq. \ref{eq:Fzzff}
can be derived for situations when (a) quadrupolar effects are
zero, (b) both first and second order quadrupolar effects are
homogeneous, (c) first order effects are inhomogeneous, while
second order effects are
homogeneous\cite{JPSJ.11.50S,Haase1993EchoS}. Thus very similar
echo decay times are expected for bulk GaAs, uniformly strained
and strain-free GaAs/AlGaAs quantum wells.

In order to characterize the timescales of the nuclear flip-flops
we introduce the flip-flop time as (Eq. 2 of the main text):
\begin{eqnarray}
T_{2,ff}=1/(T_{2}^{-1}-T_{2,zz}^{-1}),\label{eq:T2ff}
\end{eqnarray}
where $T_2$ is the experimentally measured echo decay time. When
homonuclear flip-flops are strongly suppressed, $T_{2}\rightarrow
T_{2,zz}$ according to the analysis for case (i), so that
$T_{2,ff}\rightarrow\infty$ according to Eq. \ref{eq:T2ff}. To
estimate the $T_{2,ff}$ in the absence of inhomogeneous
quadrupolar effects we use the same Eq. \ref{eq:T2ff} but with the
calculated $T_{2,zz+ff}$ instead of experimental $T_2$. The
resulting $T_{2,ff}$ values are 1.5~ms, 1.1~ms and 0.6~ms for
$^{69}$Ga, $^{71}$Ga and $^{115}$In respectively (these values are
also quoted in the main text).

We now also include the effect of heteronuclear dipole-dipole
interaction which causes additional echo decay via spectral
diffusion. This effect can not be treated exactly, but it is
possible to obtain a reasonable estimate.

First we estimate the root mean square amplitude of the effective
field that the ''fluctuating'' isotope $S$ exerts on the
''studied'' isotope $I$. For that we calculate the heteronuclear
interaction second moment:
\begin{align}
&M_{2E,S}=F\left(\frac{\mu_0}{4\pi}\right)^2\hbar^2\gamma_I^2\gamma_S^2a^{-6}\rho_S\sum_{i\ne0}\left(\frac{3}{2}\times\frac{1-3\cos^2\theta_i}{(r_i/a)^3}\right)^2\label{eq:M2Ehetero}\nonumber\\
&F=\frac{2}{9(2 S+1)},
\end{align}
where $I$ and $S$ indices denote the values of the corresponding
isotopes. In the lattice sum the summation now goes either over
the same fcc sublattice (e.g. for interaction between In and Ga),
which gives the same value as in Eq. \ref{eq:fSumSame}, or over
the ''other'' sublattice (for interaction of As with In or Ga),
which gives a lattice sum of $\approx113.30$.

The echo decay time of the $I$ isotope will depend both on the
coupling strength (characterized by $M_{2E,S}$) and the
correlation time $\tau$ of the flip-flops of the $S$ nuclei. The
fastest decoherence of the $I$ nuclei will take place when
$\tau\times \sqrt{M_{2E,S}}\sim1$. Using a Gaussian approximation
\cite{GaussApproxEchoDecS} an upper bound on the decay rate under
these conditions can be estimated as $\sim\sqrt{M_{2E,S}}/1.62$,
where the 1.62 factor is a root of a transcendental equation
derived in Ref. \cite{GaussApproxEchoDecS} under the assumption of
an exponential correlation function. The overall spin echo decay
time of the $I$ nuclei can then be calculated as
\begin{align}
T_{2,zz+ff+IS}=\left(T_{2,zz+ff}^{-1}+\sum_{S\ne
I}\sqrt{M_{2E,S}}/1.62\right)^{-1},\label{eq:T2Full}
\end{align}
where the summation goes over all isotopes $S$ distinct from $I$.
The $T_2$ times calculated according to Eq. \ref{eq:T2Full} are
given in the last row of Table \ref{tab:T2Calc} and are also shown
by the dashed lines in Fig. 3 of the main text. These estimates
are based on the maximum possible effect of the heteronuclear
coupling. However, comparing them to the $T_{2,zz+ff}$ values we
see that the heteronuclear interaction reduces $T_2$ times by no
more than $40\%$. Thus the key role in nuclear spin echo decay
(and corresponding reduction of $T_2$) is played by the
homonuclear flip-flops.

In these calculations of $T_{2,zz+ff+IS}$ we take into account
only the flip-flops of the $S_z=\pm1/2$ states and the same
chemical composition (same values of $\rho$) as for the studied
dots, which corresponds to a hypothetical quantum dot where
first-order quadrupolar shifts are strongly inhomogeneous, while
second-order shifts are homogeneous or absent. In real
self-assembled dots the first-order shifts are strongly
inhomogeneous, but the flip-flops of the $\pm1/2$ states may be
suppressed only partially and to a different degree for different
isotopes. Thus we may have an intermediate case between cases (i)
and (ii), making exact calculation of the nuclear $T_2$ difficult.
But as demonstrated above, one can obtain both the upper and lower
bounds on $T_2$: $T_{2,zz+ff+IS}<T_2<T_{2,zz}$.

The observation of $T_2\approx T_{2,zz}$ for $^{75}$As nuclei in
strained dots is then interpreted as strong suppression of the
homonuclear flip-flops due to strongly inhomogeneous second-order
quadrupolar shifts of arsenic (resulting from atomic-scale alloy
disorder). By contrast for Ga and In isotopes $T_2<T_{2,zz}$ is
observed in strained dots, implying weaker suppression of the
homonuclear flip-flop due to smaller second-order shifts.

Previously the $T_2$ values calculated using the technique of Ref.
\cite{Haase1993EchoS} were shown to deviate by less than 25\% from
the experimental values \cite{Haase1993EchoS}. As expected the
values of $T_{2,zz+ff+IS}$ calculated in the present work are in
good agreement with earlier spin-echo decay measurements on
$^{75}$As [Refs.
\cite{KondoGaAsEchoS,OnoGaAsEchoS,IshiharaGaAsEchoS}] and
$^{71}$Ga [Refs. \cite{SanadaS,MakhoninNatMatS}] in lattice
matched quantum wells and dots (see triangles in Fig. 3 of the
main text) where second-order quadrupolar effects are negligible.
Somewhat faster than predicted decay of the $^{71}$Ga echo [Refs.
\cite{SanadaS,MakhoninNatMatS}] was possibly due to the
non-selective nature of the measurements resulting in additional
echo decay due to residual first-order quadrupolar shifts.


\begin{thebibliography}{10}
\bibitem{zzEchoDecay}
The nuclear spin-echo decay due to the $\hat{I}_{z}\hat{J}_{z}$
term of the
  Hamiltonian can also be understood using the following simplified picture:
  The first $\pi/2$ rotation of the echo sequence transforms the initial $I_z$
  nuclear magnetization into $I_x$, which then evolves under the
  $\hat{I}_{z}\hat{J}_{z}$ Hamiltonian. This is equivalent to magnetization
  remaining in the $I_z$ state while the Hamiltonian is transformed into
  $\hat{I}_{x}\hat{J}_{x}$, which is a flip-flop term that does not conserve
  nuclear polarization and causes echo decay.
\end{thebibliography}

\end{document}